\documentclass{article}

\usepackage{arxiv}

\usepackage[utf8]{inputenc} 
\usepackage[T1]{fontenc}    
\usepackage{hyperref}       
\usepackage{url}            
\usepackage{booktabs}       
\usepackage{amsfonts}       
\usepackage{amsmath}
\usepackage{nicefrac}       
\usepackage{microtype}      
\usepackage{lipsum}		    
\usepackage{hyphenat}       %
\usepackage{graphicx}
\usepackage{natbib}
\usepackage{doi}
\usepackage{dcolumn}        
\usepackage{bm}             
\usepackage{array}

\title{Machine-State Embeddings as an Operational Coordinate System for Accelerator Operation}

\date{\vspace{-2em}April 5, 2026}

\author{\href{https://orcid.org/0000-0003-3814-8417}{\includegraphics[scale=0.06]{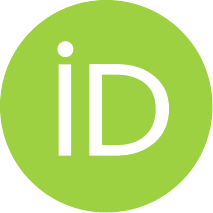}\hspace{1mm}Chris ~Tennant} \\
	Jefferson Laboratory\\
	12000 Jefferson Ave\\
	Newport News, VA 23606 \\
	\texttt{tennant@jlab.org} \\
    \\
	Jundong Li, 
    \href{https://orcid.org/0000-0003-1273-7694}{\includegraphics[scale=0.06]{orcid.pdf}\hspace{1mm}Song Wang} \\
	Department of Electrical and Computer Engineering\\
    University of Virginia\\
    351 McCormick Road \\
	Charlottesville, VA 22903 \\
}

\renewcommand{\headeright}{}
\renewcommand{\undertitle}{}

\hypersetup{
pdftitle={Machine-State Embeddings as an Operational Coordinate System for Accelerator Operation},
pdfsubject={},
pdfauthor={Chris ~Tennant},
pdfkeywords={embeddings, machine-state, particle accelerator},
}

\begin{document}
\maketitle

\begin{abstract}
We demonstrate that graph neural network (GNN) embeddings of injector configurations provide a practical operational coordinate system for the Continuous Electron Beam Accelerator Facility (CEBAF) injector at Jefferson Lab. Using 137,389 snapshots spanning January 2022 through March 2023, we show that injector operation occupies a small number of persistent, well-separated neighborhoods in a 16-dimensional learned state space rather than a featureless continuum. Density-based clustering identifies ten recurring operating regimes with strong operational run alignment, and regime persistence statistics confirm that these regimes are stable over timescales of hours to weeks. Large relocations between neighborhoods are rare and episodic; 99.6\% of one-hour operating windows fall within an empirically derived jitter baseline. Geometric outlier screening narrows a year-long dataset to a small set of intervals warranting operational review, and nearest-neighbor retrieval enables case-based reasoning over the historical archive. A controlled beam study validates that deliberate injector reconfiguration traces coherent, interpretable trajectories in embedding space. Together these capabilities demonstrate that machine-state embeddings support holistic operational monitoring in ways that single-channel inspection cannot.
\end{abstract}


\section{\label{sec:intro}Introduction}

Modern particle accelerator operation produces a machine state that is simultaneously high-dimensional and highly coupled. Magnets, RF systems, diagnostics, and beam-dependent readbacks evolve together under a web of nonlinear dependencies and hidden operational constraints. Inspecting individual signals in isolation offers an incomplete and often misleading picture of how the machine is actually behaving, and tracking the operational history of a complex beamline through time is correspondingly difficult. The fundamental challenge is representation: how should one encode the global state of a machine with hundreds of channels in a form that exposes the operational structure?

This paper explores whether a learned low-dimensional embedding can serve as a practical coordinate system for the Continuous Electron Beam Accelerator Facility (CEBAF) injector. If similar machine states map to nearby points in embedding space, then the geometry of the state space becomes a proxy for operational similarity, and a wide range of analysis tasks—identifying recurring modes, detecting unusual conditions, retrieving historical precedent—become questions about the geometry rather than about individual channels. Such an embedding does not simplify the machine by ignoring complexity; rather, it organizes that complexity into a form that can be reasoned about holistically.

The embedding analyzed here was produced by prior work in which CEBAF injector configurations were represented as heterogeneous directed graphs, with nodes corresponding to beamline elements and features encoding settings and readbacks \cite{wang2024graph}. A GNN was trained on historical data using self-supervised contrastive learning with supervised fine-tuning, yielding a 16-dimensional latent representation in which operational similarity is encoded geometrically \cite{chen2020simple}. The present study takes that embedding as its foundation and investigates whether the resulting state space supports the operational analyses listed above at scale, over a 14-month operational archive.

The contributions of this paper are threefold. First, we characterize the large-scale structure of the CEBAF injector state space and show that it consists of a small number of persistent operating regimes with strong operational run alignment and quantifiable persistence statistics. Second, we develop and validate a suite of operational analysis tools—stability assessment, outlier screening, and historical analog retrieval—that operate on the geometry of the embedding and require no hand-crafted anomaly definitions. Third, we examine where embeddings add value over single-channel monitoring and discuss the practical conditions under which alternative representations—from direct feature-space projections to transformer-based encoders—may be appropriate.

\section{\label{sec:related}Related Work}

The embedding model analyzed here builds directly on earlier work by Wang et al. \cite{wang2024graph}, in which the CEBAF injector was modeled as a heterogeneous directed graph and a GNN was trained to produce latent representations of operational configurations. That work established that the learned space encodes meaningful operational structure, motivating the present deeper analysis of what that structure implies for monitoring and operations. In the present application, the CEBAF injector was represented as a heterogeneous directed graph whose node and edge definitions were chosen to reflect the machine's operational topology and available settings and readbacks. The contrastive training objective follows Chen et al. \cite{chen2020simple}. Graph-specific extensions of contrastive learning—in which domain-appropriate augmentations (node dropping, feature perturbation) are applied to create positive pairs—have since been developed and shown to yield superior representations for graph-structured data \cite{You2020graphcl}.

Machine learning has been identified as a high-priority opportunity across virtually all domains of accelerator science—from beam diagnostics and fault detection to tuning and surrogate modeling \cite{02_Edelen2024}. In the broader context of accelerator physics, the problem of making sense of high-dimensional diagnostic data has been approached in several ways. At the LHC, Fol et al. \cite{Fol2020, fol2021beam} applied unsupervised learning to beam diagnostic data for BPM fault detection using Isolation Forest \cite{liu2008isolation}—a method closely related to the k-nearest-neighbor (kNN) outlier scoring used in Section \ref{sec:utility} below—and demonstrated that learned representations can identify faulty instruments and reveal patterns in operational behavior without requiring labeled examples. Condition monitoring in industrial machinery has seen analogous development. Wójcik and Barszcz \cite{wojcik2024condition} showed that training a variational autoencoder \cite{Kingma2014} on healthy data alone produces a latent space whose geometry supports anomaly detection through distance-based measures, without requiring any labeled fault data—a paradigm that underlies the present work as well. Once a meaningful latent representation has been learned, a natural next question is how to inspect and interpret its geometry. In that context, principal component analysis (PCA) \cite{jolliffe2002pca} and the broader toolkit of nonlinear dimensionality reduction---including t-SNE \cite{vanderMaaten2008} and UMAP \cite{McInnes2018}---have established that low-dimensional projections can expose cluster structure, transitional states, and outliers in high-dimensional scientific datasets, motivating their use as diagnostic tools across domains. The insight that low-dimensional projections of complex process data can reveal actionable operational structure has also been demonstrated in chemical process control by Joswiak et al. \cite{joswiak2019chem}, who benchmarked dimensionality reduction methods for visualizing high-dimensional plant data and showed that projections expose operating regimes, transitions, and outliers even in strongly nonlinear, highly coupled processes. The CEBAF injector shares those properties, and the results below can be understood as an accelerator-specific realization of this broader paradigm.

\section{\label{sec:embedding}Embedding and Dataset}

\subsection{\label{sec:gnn}Graph Neural Network Embedding}

Each injector snapshot is represented as a heterogeneous directed graph in which nodes correspond to instrumented beamline elements (correctors, quadrupoles, solenoids, SRF and warm RF cavities, BPMs, BLMs, ion pumps, and current monitors) and directed edges encode the physical relationships among them. Each node carries a feature vector encoding the settings and readbacks of the corresponding element. A GNN processes this graph representation and maps it to a 16-dimensional embedding vector. The GNN was trained using a self-supervised contrastive objective on unlabeled data \cite{wang2024graph} \cite{chen2020simple}. The training procedure encourages similar operational configurations to map to nearby points in the embedding space, while dissimilar configurations are pushed apart. The resulting embedding is not a compression of individual channel values but an encoding of the relational structure of the whole machine state.

For visualization, PCA is applied to the 16-dimensional embeddings to obtain a 2D projection. Except where explicitly noted, all quantitative analyses in this paper—distances, clustering, outlier scoring, and retrieval—are performed in the full 16-dimensional space. PCA serves solely as a tool for human-interpretable visualization of results computed in the higher-dimensional space.

\subsection{\label{sec:data}Dataset}

The dataset consists of 137,389 injector snapshots spanning January 4, 2022 through March 20, 2023, covering 439 days of operation. Snapshots were collected every two minutes whenever the injector beam current was at least $0.05~\mu\text{A}$. Beam current across the dataset ranged from 0.05 to $151.37~\mu\text{A}$. Each snapshot carries a 16-dimensional GNN embedding, 126 setting features, 267 reading features, and a 393-dimensional normalized feature vector constructed from the full machine state. Index alignment and timestamp consistency were applied as preprocessing steps to ensure that all retained records represent valid, consistent injector configurations.

\section{\label{sec:statespace}Injector Operations as a Structured State-Space Landscape}

\begin{figure}
\centering
\includegraphics[width=0.85\textwidth]{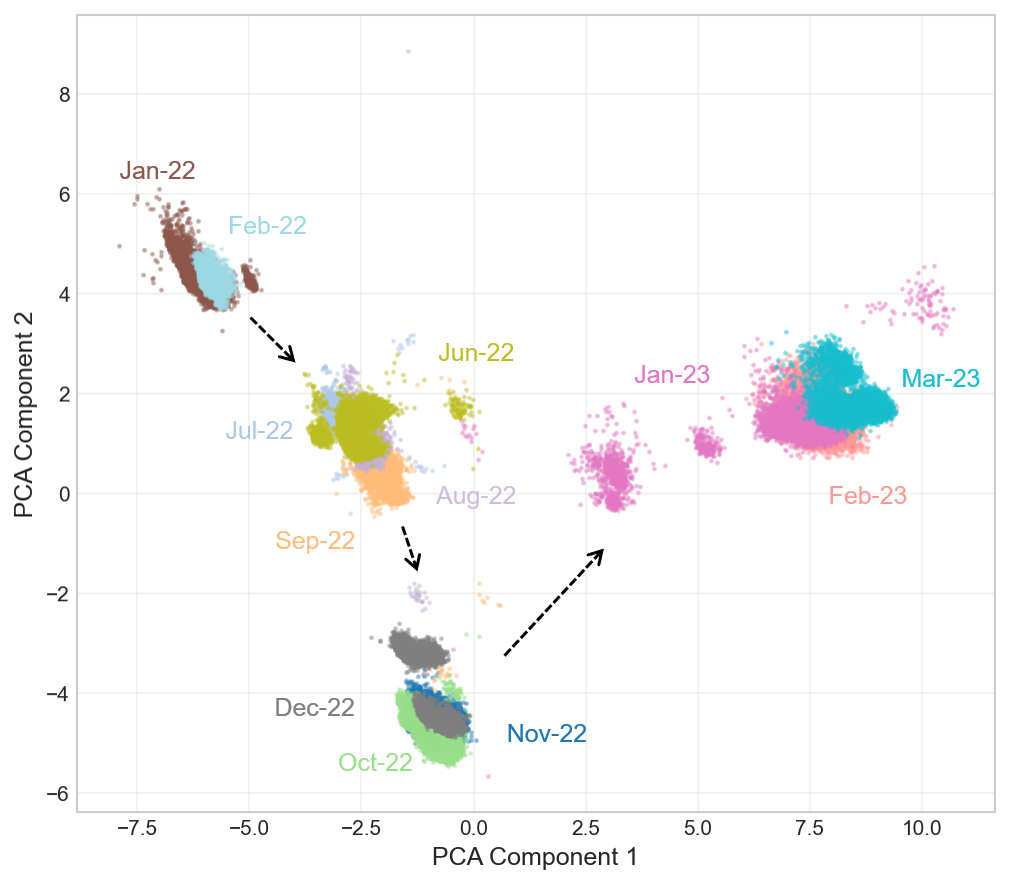}
\caption{\label{fig:monthpca}PCA projection of the GNN embedding colored by 
calendar month. The operational history from January 2022 through March 2023 
occupies several discrete islands rather than a single contiguous cloud. Dashed 
arrows indicate the general progression of time.}
\end{figure}

\begin{figure}
\centering
\includegraphics[width=\textwidth]{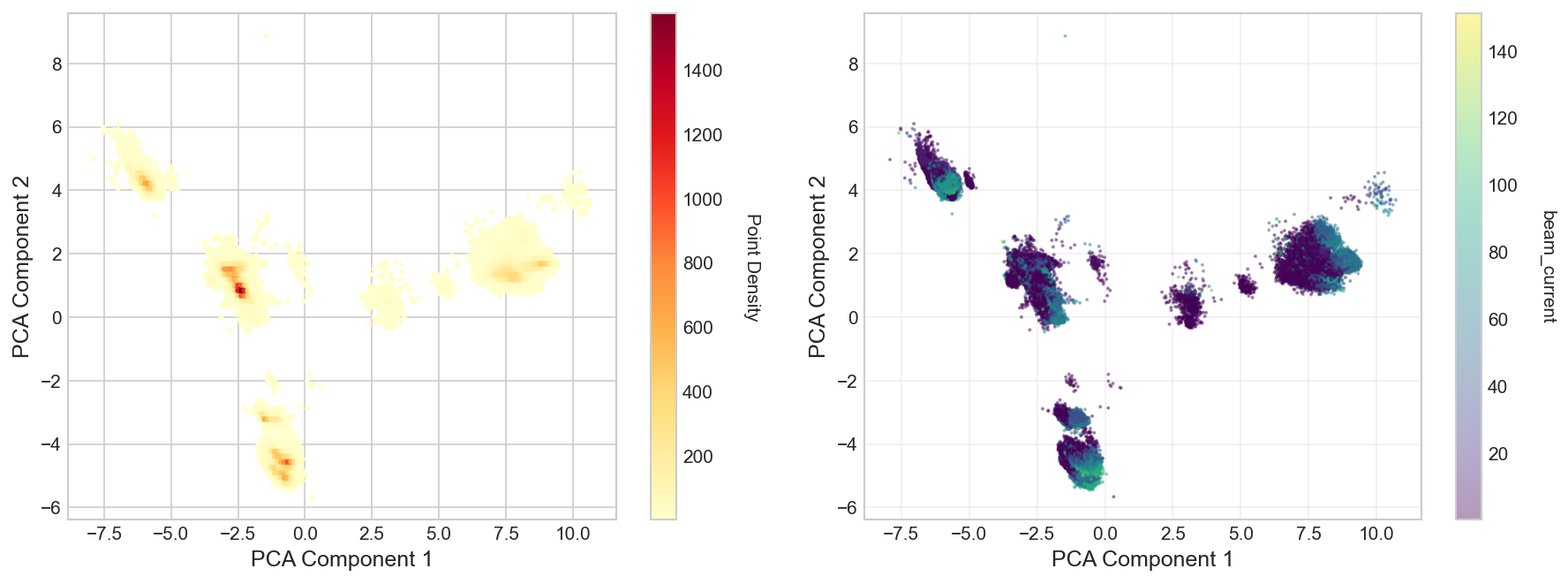}
\caption{\label{fig:densitycurrent}PCA projection colored by point density 
(left) and beam current (right). Dense cores mark frequently occupied operating 
neighborhoods; beam current varies within neighborhoods rather than aligning 
with a single global direction, indicating that embedding structure reflects 
more than current alone.}
\end{figure}

A natural first question is whether the embedding organizes injector history into interpretable structure or merely redistributes noise in a lower-dimensional space. The month-colored PCA projection in Fig.~\ref{fig:monthpca} provides an immediate answer: the 14-month operational history does not fill the 2D plane uniformly but instead occupies several clearly separated islands. The January--February 2022 data concentrate in a compact region in the upper-left. The mid-2022 datasets (June--August) populate a central-left region with visible internal substructure. The late-2022 data form a separate lower region, and early-2023 data cluster predominantly in the right-hand portion. These islands are not mere visualization artifacts; they are consistent across the density and beam-current views shown in Fig. \ref{fig:densitycurrent}. Each island contains one or more high-occupancy cores—the most frequently visited operating neighborhoods—surrounded by lower-density halos and sparse satellite patches consistent with brief excursions or transitional states. Beam current varies within these regions rather than mapping one-to-one onto a single global direction, implying that the embedding encodes a richer operational signature than beam current alone could provide.

\subsection{\label{sec:recurring}Recurring Operating Regimes}

To move from a visual impression to a quantifiable characterization, we applied HDBSCAN \cite{campello2013hdbscan, campello2015hierarchical} directly to the 16-dimensional embedding space. HDBSCAN is a density-based clustering algorithm that identifies groups of points in dense regions while leaving isolated or weakly connected points unlabeled as noise. It does not require the number of clusters to be specified in advance and is therefore well suited to an embedding space that may contain groups of different size and density.

\begin{figure}
\centering
\includegraphics[width=0.85\textwidth]{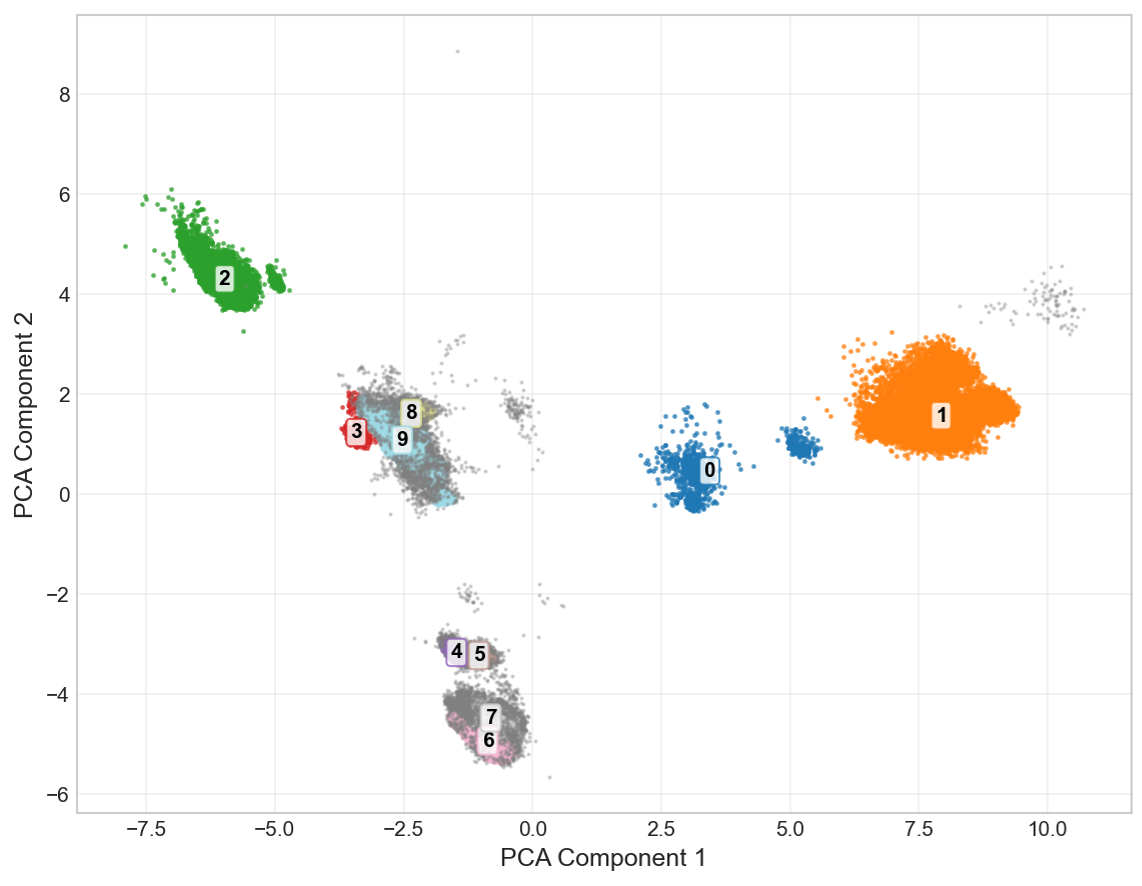}
\caption{\label{fig:clustermap}HDBSCAN clusters identified in the 16D embedding 
space and visualized in the 2D PCA projection. Cluster labels mark recurring 
operating regimes and gray points denote noise or weakly assigned states.}
\end{figure}

The analysis identifies ten recurring operating regimes and a noise class. Only 4,694 points (3.4\%) are classified as noise, indicating that the great majority of operational history falls cleanly within one of the identified regimes. These ten clusters are best understood as a descriptive summary of recurring operational modes in this dataset. The annotated cluster map is shown in Fig. \ref{fig:clustermap}.

A small number of clusters dominate dwell time. Cluster 9 is the largest (43,094 points, mean beam current $30.96~\mu\text{A}$), followed by Cluster 1 (31,616 points, $42.44~\mu\text{A}$), Cluster 7 (18,216 points, $50.53~\mu\text{A}$), Cluster 2 (16,142 points, $38.16~\mu\text{A}$), and Cluster 6 (12,816 points, $40.21~\mu\text{A}$). Together these five clusters account for the large majority of operational time.

\subsection{\label{sec:persistence}Regime Persistence}

That clusters exist does not by itself establish that they are persistent—they could in principle correspond to frequently revisited but short-lived excursions. To quantify persistence, we computed, for each cluster, the number of times the machine entered it (defined at the 8-hour shift level), the median contiguous dwell time in hours, and the 90th-percentile dwell time. Results are summarized in Table \ref{tab:persistence}. A dwell run is a maximal sequence of consecutive shifts in the same cluster; a run ends when the cluster assignment changes or when one or more shifts are absent from the data (i.e., a gap of at least 8 hours with no samples).

\begin{table}
\centering
\caption{\label{tab:persistence}Summary of operating regimes identified by HDBSCAN
clustering in the 16D embedding space. Each 8-hour shift is assigned its dominant 
cluster (by mode). Persistence statistics are computed across all such runs. Rows 
are sorted by descending median dwell time.}
\begin{tabular}{cccc}
\toprule
Cluster ID & No.\ Entries & Median Dwell (hrs) & $p_{90}$ Dwell (hrs) \\
\midrule
1   &  4 & 396 & 581.6 \\
7   &  3 & 160 & 448.0 \\
9   & 15 & 136 & 275.2 \\
6   &  5 & 104 & 236.8 \\
2   &  7 &  88 & 280.0 \\
0   &  1 &  88 &  88.0 \\
8   &  3 &  64 &  83.2 \\
5   &  2 &  44 &  66.4 \\
3   &  2 &  24 &  30.4 \\
4   &  4 &  16 & 116.8 \\
$-1$ & 24 &   8 &  13.6 \\
\bottomrule
\end{tabular}
\end{table}

\begin{figure}
\centering
\includegraphics[width=1.0\textwidth]{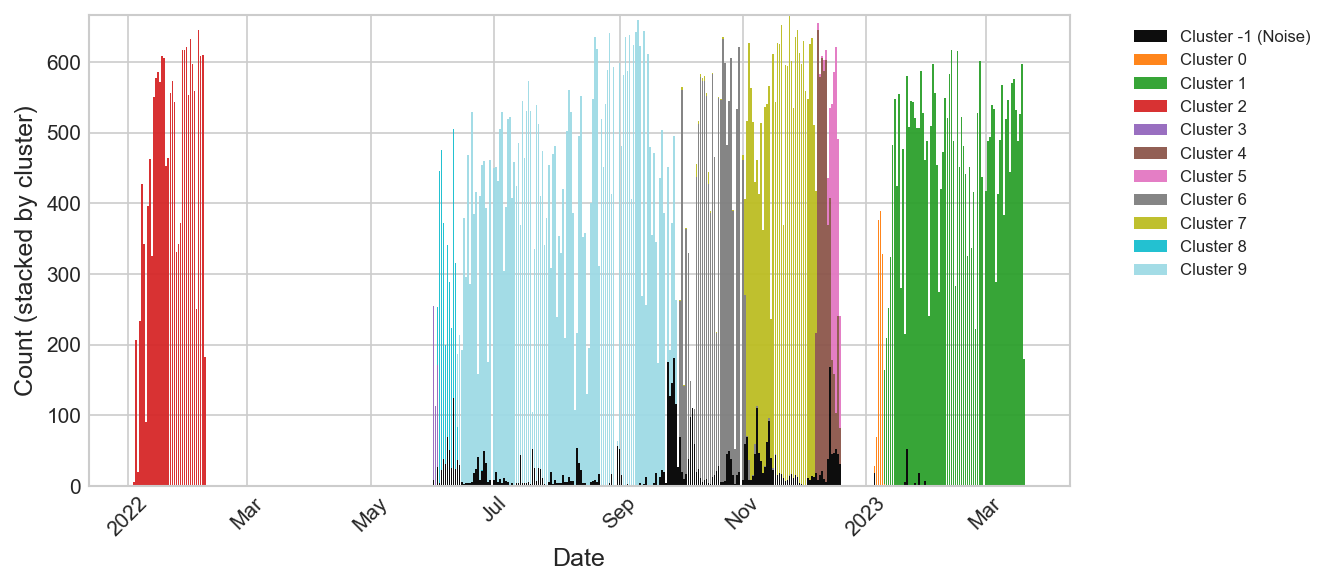}
\caption{\label{fig:stackedhist}Time distribution of cluster membership shown 
as a stacked histogram. Long contiguous blocks indicate that operating regimes 
persist over extended periods rather than being uniformly interleaved in time.}
\end{figure}

The dominant clusters are entered only a handful of times but maintain median dwells measured in days: Cluster 1 entered four times with a median dwell of 396 hours ($\sim$16.5 days) and a 90th-percentile dwell of 581.6 hours ($\sim$24 days). Cluster 7 entered three times with a median of 160 hours ($\sim$6.7 days). This persistence is also visible in the stacked timestamp histogram of Fig. \ref{fig:stackedhist}, which shows clusters dominating long, contiguous run periods rather than being uniformly interleaved in time. The noise class ($-$1), by contrast, enters 24 times with a median dwell of only 8 hours, confirming that unclassified periods are typically transitional episodes rather than an alternative stable operating mode.

One notable feature of Fig. \ref{fig:stackedhist} is that the noise label is absent during January–February 2022. A plausible interpretation is that early-2022 operation was dominated by a single especially dense and homogeneous operating region, with all nearby states absorbed into clusters. Beginning in June 2022, the embedding space becomes more fragmented as distinct operating modes multiply, creating sparse inter-cluster regions that HDBSCAN naturally assigns to noise.

\subsection{\label{sec:fingerprints}Cluster Fingerprints}

Geometric regimes are operationally useful only if they connect back to the underlying machine. For each cluster, we compute a fingerprint by comparing that cluster's mean settings and readings to the corresponding full-dataset means. The top-ranked features by absolute delta are those that most strongly distinguish the regime from typical operation. To illustrate the interpretive value of this approach, we focus on two clusters whose fingerprints map cleanly onto recognizable injector subsystems: Cluster~2 and Cluster~6.

Table~\ref{tab:fingerprints} lists the top four settings and top four readings for each of these two clusters. Cluster~2 is defined primarily by four warm chopper cavities—CHOP1Y and CHOP1X (the vertical and horizontal settings of the first chopper), and CHOP2X and CHOP2Y (the corresponding settings of the second chopper). The choppers control the pulse structure of the beam extracted from the electron gun, and their dominance in the fingerprint indicates that Cluster~2 corresponds to a mode defined by a specific configuration during the beam formation stage.

Cluster~6, by contrast, is defined by four SRF cavity setpoints in the 0L04 cryomodule—the last module in the injector before the beam enters the main linac. Specifically, 0L04-8, 0L04-6, 0L04-5, and 0L04-3 are individual cavities within that module. The associated readings are ion pump currents and beam position monitors along the medium-energy transport line. Together, the Cluster~6 fingerprint points to a mode in which the energy delivered by the final injector cryomodule is the primary organizing variable, distinguishing it geometrically from other clusters that share similar chopper or gun settings but differ in 0L04 configuration.

These two examples illustrate the broader point: fingerprints do not establish causality, but they do provide an interpretable bridge from abstract geometric structure back to specific subsystems, enabling an operator or subject matter expert to reason about what physical configuration each recurring regime represents.

\begin{table}
\centering
\caption{\label{tab:fingerprints}Top four setting and reading features 
distinguishing Cluster~2 and Cluster~6 from the full-dataset average, 
based on largest absolute cluster-minus-global-mean deltas. Element types 
are noted in parentheses.}
\begin{tabular}{lll}
\toprule
 & Top 4 Settings & Top 4 Readings \\
\midrule
Cluster 2 & CHOP1Y (chopper 1 - vertical), & VIP0L0450 (ion pump), \\
          & CHOP1X (chopper 1 - horizontal), & IPM0L10 (beam position monitor), \\
          & CHOP2X (chopper 2 - horizontal), & VIP0L04B (ion pump), \\
          & CHOP2Y (chopper 2 - vertical) & VIP0L05A (ion pump) \\[4pt]
Cluster 6 & 0L04-8 (SRF cavity), & VIP0L00 (ion pump), \\
          & 0L04-6 (SRF cavity), & IPM0R04 (beam position monitor), \\
          & 0L04-5 (SRF cavity), & IPM0L02 (beam position monitor), \\
          & 0L04-3 (SRF cavity) & IPM0L04 (beam position monitor) \\
\bottomrule
\end{tabular}
\end{table}

\section{\label{sec:trajectory}Dynamics of Motion Through State-Space}

Having established that the embedding is organized into persistent neighborhoods, we turn to how the machine moves through that space over time. The analysis reveals a characteristic pattern across multiple timescales: most motion is small and routine, while large relocations are rare and episodic.

\subsection{\label{sec:step}Shift-to-Shift Step Sizes}

\begin{figure}
\centering
\includegraphics[width=0.75\textwidth]{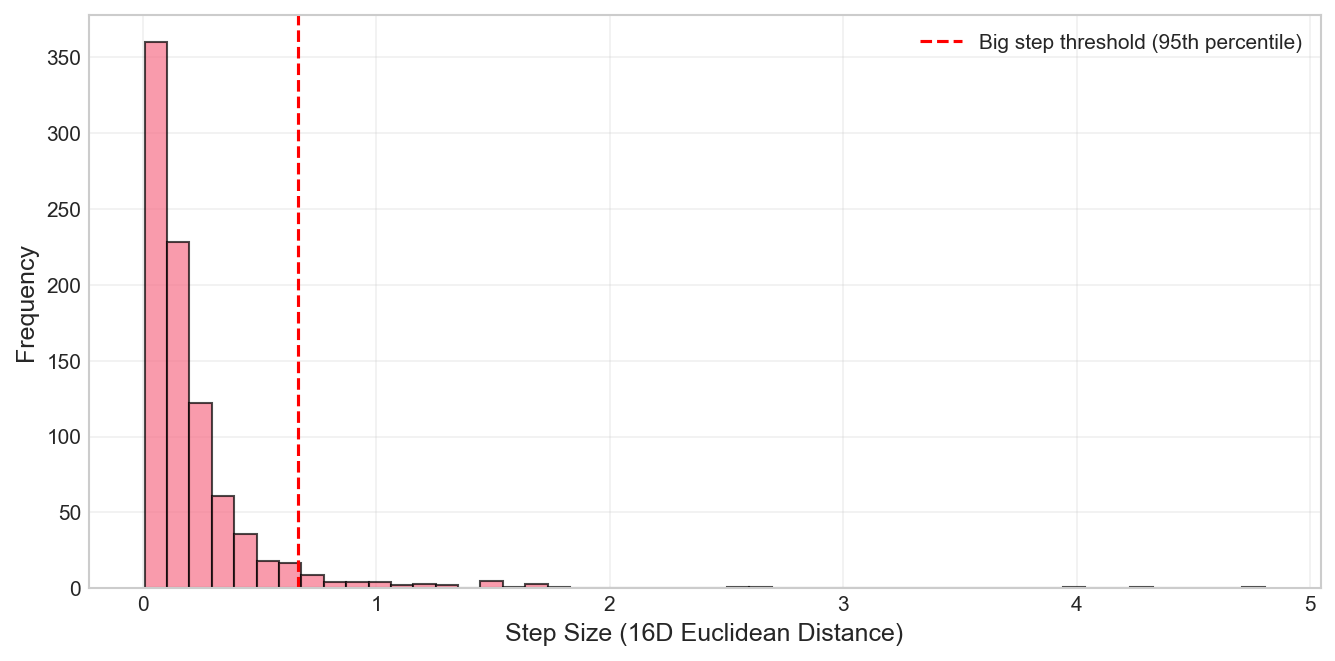}
\caption{\label{fig:stepsize}Distribution of step sizes for the 884 
shift-to-shift transitions, computed as the Euclidean distance between 
consecutive shift centroids in 16D embedding space. The distribution is strongly 
right-skewed; the dashed line marks the 95th-percentile ``big step'' threshold 
of 0.67.}
\end{figure}

To quantify motion at an operationally meaningful timescale, we aggregate snapshots into owl, day, and swing shifts and represent each shift by its centroid in the 16-dimensional embedding space. The Euclidean distance between consecutive shift centroids then measures shift-to-shift reconfiguration. Across 885 shifts (884 transitions), the step-size distribution is strongly right-skewed. The distribution peaks sharply near zero, with most transitions clustered well below 0.3 in 16D Euclidean distance, while a long tail of rare large moves extends toward values above 1.0 (Fig. \ref{fig:stepsize}).

\begin{figure}
\includegraphics[width=\columnwidth]{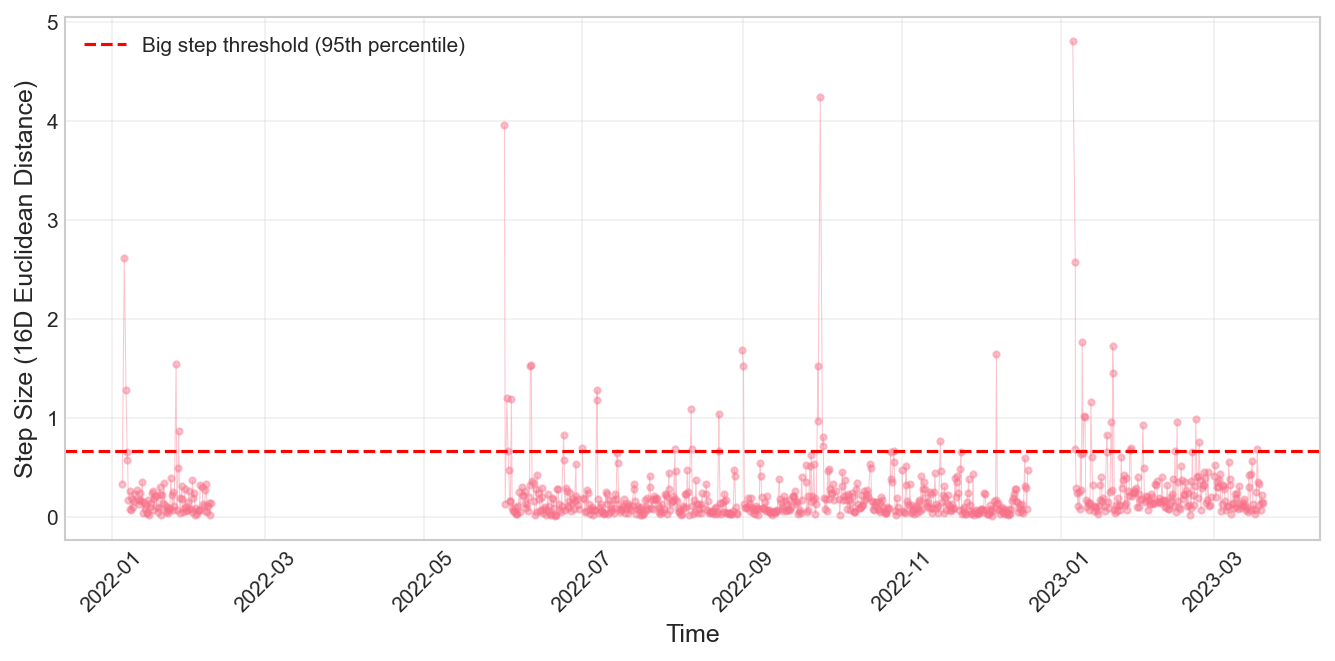}
\caption{\label{fig:steptime}Shift-to-shift embedding-space step sizes as a 
function of time. The dashed line marks the 95th-percentile threshold; large 
relocations appear as isolated spikes rather than sustained elevations.}
\end{figure}

Using the 95th-percentile value of 0.67 as a threshold for large relocations, these events appear as isolated spikes in time (Fig. \ref{fig:steptime}) rather than sustained elevations. The largest jumps cluster on a few specific dates: January 5, 2023, September 30, 2022, and June 1, 2022. Table \ref{tab:stepsize} lists the ten largest transitions and their associated beam current changes. Several large steps correspond to returns from extended downtime.

\begin{table}
\centering
\caption{\label{tab:stepsize}The ten largest shift-to-shift step sizes, with 
associated beam-current change and gap since the previous shift. Step size is 
the Euclidean distance between consecutive shift centroids in the 16D embedding 
space.}
\begin{tabular}{l D{.}{.}{2} D{.}{.}{2} D{.}{.}{2}}
\toprule
Date--Time & \multicolumn{1}{c}{Step Size} & \multicolumn{1}{c}{$\Delta I$ ($\mu$A)} 
           & \multicolumn{1}{c}{Gap (days)} \\
\midrule
2023-01-05 16:04 & 4.81 & -0.57  &  17.30 \\
2022-09-30 13:10 & 4.24 & -6.57  &   0.58 \\
2022-06-01 00:50 & 3.95 & -24.25 & 112.73 \\
2022-01-05 18:14 & 2.62 &  18.86 &   0.47 \\
2023-01-06 15:48 & 2.57 &  33.36 &   0.83 \\
2023-01-09 08:28 & 1.77 & -0.12  &   0.02 \\
2023-01-21 00:16 & 1.72 &  23.59 &   0.01 \\
2022-08-31 12:30 & 1.69 & -26.99 &   1.52 \\
2022-12-07 08:00 & 1.65 & -48.40 &   0.00 \\
2022-01-25 17:08 & 1.54 & -33.89 &   0.40 \\
\bottomrule
\end{tabular}
\end{table}

Within-shift variability also differs by shift type. Mean within-shift path lengths are 32.28 for day shifts, 36.96 for swing shifts, and 41.28 for owl shifts, suggesting that day operations are typically more stable. When examining the shift handoff—the cumulative embedding-space step associated with entering each shift type, excluding long-downtime gaps—the smoothest routine handoff is swing-to-owl (cumulative step 54.33), whereas the largest reconfiguration tends to accompany the owl-to-day transition (67.91).

\subsection{\label{sec:drift}Long-Horizon Drift}

\begin{figure}
\centering
\includegraphics[width=0.75\textwidth]{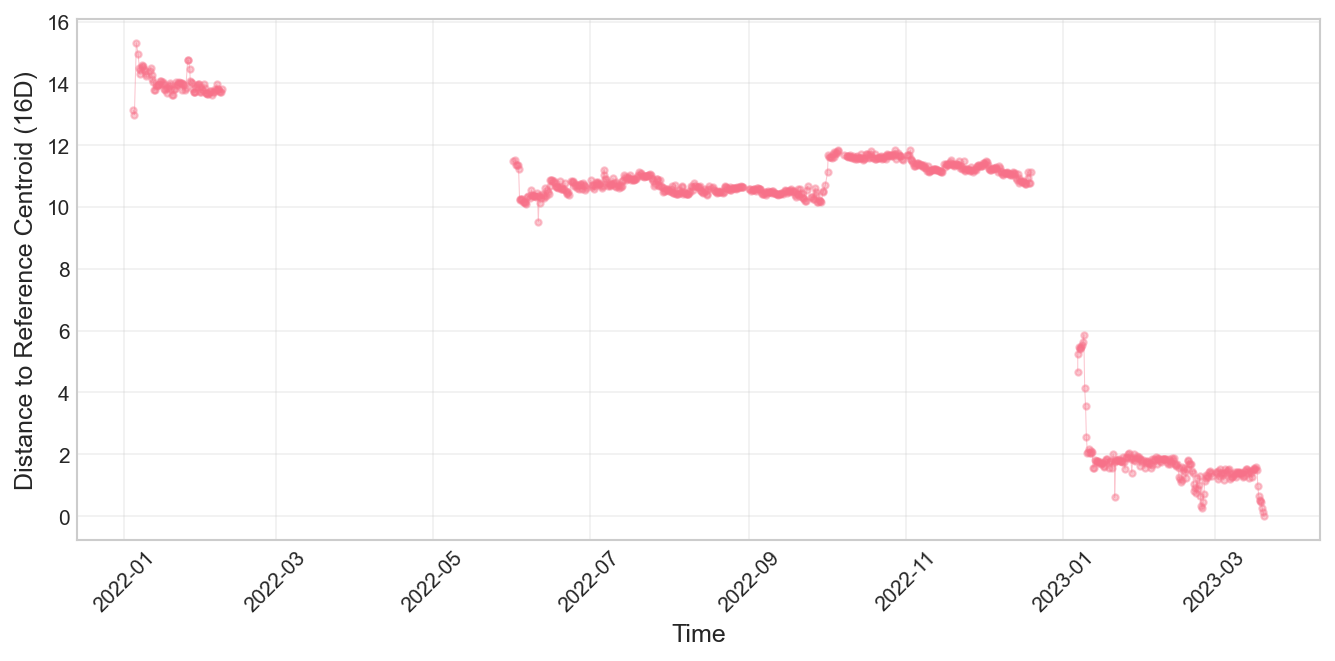}
\caption{\label{fig:drift}Distance of each shift centroid from the final shift 
(March 20, 2023), plotted over time. Breaks indicate extended beam-off periods. 
The structure of stable plateaus and abrupt jumps reinforces the picture of 
persistent regimes with rare large transitions.}
\end{figure}

The drift-to-reference metric measures the distance of each shift centroid from the final shift of the dataset (March 20, 2023). As shown in Fig. \ref{fig:drift}, this view reinforces the picture of discrete relocations. Early 2022 lies far from the final reference, mid- to late-2022 occupies an intermediate plateau with a visible jump around early October, and early 2023 collapses to much smaller distances after the abrupt January transition. The machine spends long stretches at approximately constant distance from the reference—confirming extended residence in stable neighborhoods—punctuated by rapid steps.

\subsection{\label{sec:sept}The September 30, 2022 Case Study}

Among the large relocation events, the September 30, 2022 transition (step size 4.24, second largest in the dataset) provides an opportunity to connect embedding-space motion to specific operational context. To investigate, logbook entries from September 28–October 1, 2022 were extracted, filtered for human-authored narrative content, and analyzed using large language models. The logbook analysis workflow is described in the Section~\ref{sec:supplement}.

Both models converge on a compound operational event rather than a single isolated cause. A full maintenance day on September 29 was followed by RF recovery and setup changes intended to support multi-hall running, including gangphase and path-length adjustments. Recovery was then disrupted by cascading faults: a North Linac 2K cold-box trip, cavity instabilities and de-ratings, magnet overtemperature faults, and additional control issues. The resulting large embedding-space jump reflects not a single parameter change but a substantial, partially degraded reconfiguration of the injector following maintenance and multi-system recovery. This is precisely the kind of event that is difficult to detect from any single channel but is visible as a large step in the holistic machine-state representation.

\subsection{\label{sec:transition}Regime Transition Dynamics}

Downsampling to the shift level, the regime-to-regime transition graph reduces to a sparse set of pathways. Across the full run there are 50 shift-scale transitions spanning 19 unique directed pairs. The most common exchanges involve the noise label: transitions from noise ($-$1) to Cluster 9 and the reverse each occur 12 times, while other pairings occur only a handful of times (Table \ref{tab:transitions}). This sparsity confirms that the machine rarely changes regimes at the shift scale. When it does, the change often passes through transitional noise states rather than jumping directly between clusters.

\begin{table}
\centering
\caption{\label{tab:transitions}Most frequent regime transitions at the shift 
level. The dominance of noise-mediated transitions indicates that inter-regime 
moves typically pass through unclassified states.}
\begin{tabular}{ccc}
\toprule
From Cluster & To Cluster & Count \\
\midrule
$-1$ & 9    & 12 \\
9    & $-1$ & 12 \\
8    & $-1$ &  3 \\
$-1$ & 8    &  2 \\
\bottomrule
\end{tabular}
\end{table}

When the machine enters the noise label, it typically does not remain there long: 651 separate noise runs are identified, with a median dwell time of 8.0 minutes and a mean of 14.8 minutes, though the distribution is heavy-tailed with a maximum of 268 minutes. Note that the fifth longest noise dwell occurs in late September (Table \ref{tab:noisedwell}), which aligns with the degraded machine state documented in the September 30 case study above.

\begin{table}
\centering
\caption{\label{tab:noisedwell}Five longest contiguous dwell periods in the 
HDBSCAN noise class ($-1$). The concentration in late September--early October 
2022 is consistent with the compound operational event documented in the 
September~30 case study.}
\begin{tabular}{llcc}
\toprule
Start & End & No.\ Points & Dwell (min) \\
\midrule
10/06/2022 17:06 & 10/06/2022 21:34 &  95 & 268 \\
10/07/2022 21:28 & 10/08/2022 01:04 & 105 & 216 \\
12/14/2022 06:10 & 12/14/2022 09:06 &  62 & 176 \\
06/11/2022 01:06 & 06/11/2022 03:50 &  72 & 164 \\
09/28/2022 01:30 & 09/28/2022 04:10 &  48 & 160 \\
\bottomrule
\end{tabular}
\end{table}

\section{\label{sec:utility}Operational Utility}

\subsection{\label{sec:stability}Stability Baseline and Anomaly Screening}

Identifying genuinely unusual intervals requires an empirical baseline for expected variability under stable operation. We define this baseline using anchor windows---multi-hour periods during which settings remained essentially constant---computing the $L_2$ reading-space dispersion within each window as a reference distribution. Concretely, for a window containing $N$ snapshots with reading vectors $\mathbf{r}_1, \ldots, \mathbf{r}_N \in \mathbb{R}^d$, the dispersion is defined as the mean pairwise Euclidean distance from each point to the window centroid $\bar{\mathbf{r}}$:
\begin{equation}
    D = \frac{1}{N} \sum_{i=1}^{N} \left\| \mathbf{r}_i - \bar{\mathbf{r}} \right\|_2.
    \label{eq:dispersion}
\end{equation}
Five anchor windows were identified (901 samples total) and their pooled 95th-percentile dispersion of 12.44 defines the jitter baseline. The stability ratio for any one-hour operating window is then
\begin{equation}
    s = D_\text{window} / D_{p95},
    \label{eq:stability}
\end{equation}
where $D_\text{window}$ is the reading-space dispersion for that window and $D_{p95}$ is the pooled baseline. Values of $s \leq 1$ are consistent with routine jitter whereas values above 1 indicate dispersion exceeding the empirical noise floor.

\begin{figure}
\centering
\includegraphics[width=0.85\textwidth]{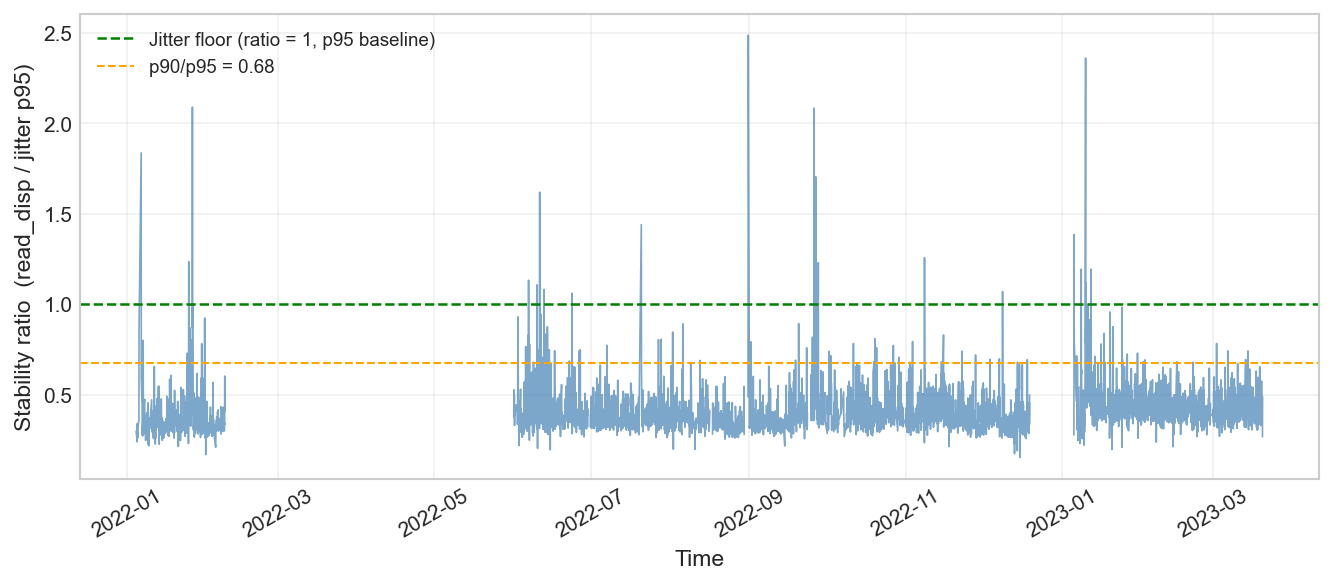}
\caption{\label{fig:stabilitytime}Stability ratio (reading-space dispersion 
normalized to the pooled anchor-window $p_{95}$ baseline) for all one-hour 
windows across the 14-month dataset. Only 22 of 6{,}263 windows (0.4\%) exceed 
the jitter threshold ($s = 1$, solid line).}
\end{figure}

\begin{figure}
\centering
\includegraphics[width=0.85\textwidth]{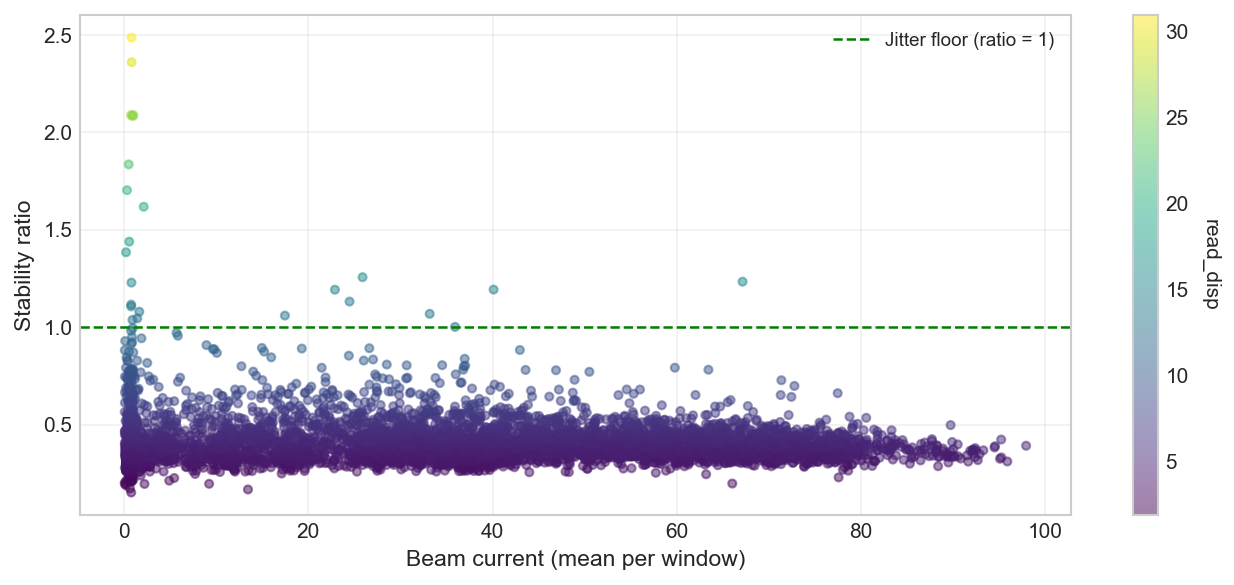}
\caption{\label{fig:stabilitycurrent}Stability ratio versus mean beam current 
per one-hour window. Elevated ratios occur primarily at very low current, 
consistent with setup or recovery activity rather than instability during normal 
beam delivery.}
\end{figure}

Of the 6,263 one-hour windows evaluated across the full dataset, 6,241 (99.6\%) have stability ratio $s \leq 1$, with a median of 0.38. Only 22 windows (0.4\%) exceed the jitter baseline, appearing as isolated spikes (Fig. \ref{fig:stabilitytime}) rather than sustained periods of elevated variability. The maximum ratio observed is 2.36. As shown in Fig. \ref{fig:stabilitycurrent}, elevated stability ratios are concentrated at very low beam current, whereas moderate- and high-current operation clusters uniformly below $s = 1$. This pattern suggests that departures from the baseline are associated with low-current setup or recovery periods rather than with instability during established beam delivery.

\subsection{\label{sec:outlier}Geometric Outlier Screening}

\begin{figure}
\centering
\includegraphics[width=0.85\textwidth]{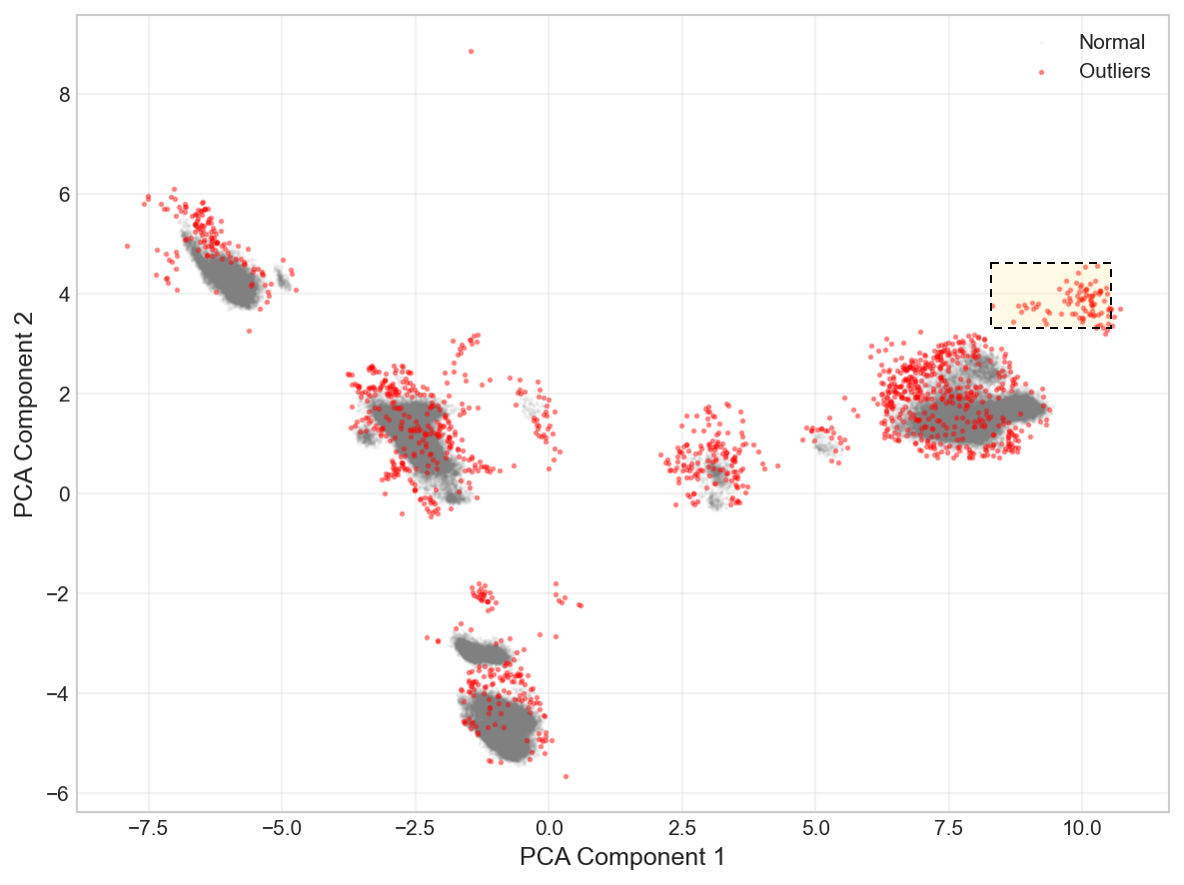}
\caption{\label{fig:knnpca}Outliers (red) overlaid on the PCA projection. 
Flagged points lie preferentially at regime boundaries and in sparse inter-island 
regions. The dashed box marks the area associated with the longest outlier 
interval (January 21, 2023).}
\end{figure}

In the PCA overlay (Fig. \ref{fig:knnpca}), flagged points concentrate near the boundaries of major operating islands and in low-density inter-cluster regions, indicating that outliers are typically excursions away from dominant neighborhoods rather than evidence for additional undetected regimes. Merging temporally adjacent outliers with a 300-second gap threshold yields 841 distinct intervals, confirming that anomalous episodes are usually brief and fragmented. The three longest intervals are: 5,640 s on January 21, 2023 (01:28–03:02), 2,520 s on July 20, 2022 (20:24–21:06), and 2,400 s on January 27, 2023 (20:18–20:58).


The two screening approaches—noise-dwell labeling and kNN-based isolation scoring—capture related but distinct notions of unusual behavior. HDBSCAN labels points as noise when they fall in regions that are too sparse to be assigned confidently to any of the identified clusters. The kNN analysis, by contrast, is not a classification step. For each point, it computes a distance-based isolation score from its local neighborhood in the full 16D embedding space, using the distances to its $k$ nearest neighbors \cite{ramaswamy2000efficient, breunig2000lof}. Larger scores indicate that the point is geometrically more isolated from the historical cloud of operating states. Table \ref{tab:comparison} summarizes the key distinctions. Cross-referencing the two provides a stronger operational signal, a sustained noise dwell that also contains many high-scoring kNN outliers is more likely to reflect a genuine machine excursion than either indicator alone.

\begin{table}
\centering
\caption{\label{tab:comparison}Comparison of the two anomaly-related analyses. 
The methods are complementary: noise dwell asks how long the machine remained 
outside the main regimes while kNN scoring asks how geometrically isolated each 
state was from the full operational history. For the kNN-based interval analysis, 
temporally adjacent outlier points were merged into a single interval when they 
were separated by no more than 300~s; this is the ``merge gap'' referred to in 
the text.}
\begin{tabular}{lll}
\toprule
 & Noise Cluster ($-1$) & kNN Outliers \\
\midrule
Source       & HDBSCAN density clustering      & kNN mean distance score \\
Scope        & Low-density unclassified points & All 137{,}389 points scored \\
Threshold    & HDBSCAN internal density        & 99th percentile of kNN score \\
Count        & 4{,}694 points (3.4\%)          & 1{,}374 points (1.0\%) \\
Merge gap    & 10-min gap                      & 5-min (300\,s) merge gap \\
\bottomrule
\end{tabular}
\end{table}

To test whether combining the two methods identifies operationally meaningful intervals, we computed the overlap between the ten longest noise dwells and the kNN outlier intervals. Results are presented in Table \ref{tab:overlap}. The overlap is generally sparse with most of the longest noise dwells have little or no concurrent kNN outlier activity. One case stands out—the January 21, 2023 dwell of 94 minutes overlaps with outlier intervals at 100\%. This episode is worth examining in detail precisely because it demonstrates the value of the combined screen in an unexpected way. Manual inspection revealed that multiple beam position monitors and corrector setpoints returned undefined values in the archiver during this period, pointing to a data-quality failure rather than a beam-physics event. This outcome illustrates its utility. The two independent geometric screens—one based on cluster membership, the other on nearest-neighbor isolation—converged on the same 94-minute window without any prior knowledge of the archiver fault. A screening tool that surfaces both unusual machine states \emph{and} unusual data states is operationally valuable, since corrupted inputs to an embedding model can produce misleading embeddings regardless of the underlying machine behavior. The January 21 case therefore motivates a practical refinement: applying data-quality masks prior to embedding or anomaly scoring, i.e., flagging and excluding samples or channels with undefined, missing, or otherwise invalid archived values so that instrumentation failures are not misinterpreted as genuine operational excursions.

\begin{table}
\centering
\caption{\label{tab:overlap}Overlap analysis between the ten longest HDBSCAN 
noise dwells and kNN outlier intervals. Most long noise dwells have little or 
no concurrent outlier activity. The January~21, 2023 entry (marked with an 
asterisk) shows 100\% overlap and was traced to an archiver data-quality 
failure affecting multiple process variables.}
\begin{tabular}{llcccc}
\toprule
Noise Start       & Noise End         & Dwell (min) & Overlap? & Overlap (min) & Overlap (\%) \\
\midrule
10/06/2022 17:06  & 10/06/2022 21:34  & 268 & No  &  0 &   0   \\
10/07/2022 21:28  & 10/08/2022 01:04  & 216 & No  &  0 &   0   \\
12/14/2022 06:10  & 12/14/2022 09:06  & 176 & No  &  0 &   0   \\
06/11/2022 01:06  & 06/11/2022 03:50  & 164 & Yes &  8 &   4.9 \\
09/28/2022 01:30  & 09/28/2022 04:10  & 160 & No  &  0 &   0   \\
08/11/2022 22:10  & 08/12/2022 00:32  & 142 & Yes &  6 &   4.2 \\
12/14/2022 02:04  & 12/14/2022 03:52  & 108 & No  &  0 &   0   \\
11/07/2022 10:24  & 11/07/2022 12:04  & 100 & Yes & 10 &  10.0 \\
11/03/2022 16:04  & 11/03/2022 17:38  &  94 & No  &  0 &   0   \\
01/21/2023 01:28* & 01/21/2023 03:02* &  94 & Yes & 94 & 100   \\
\bottomrule
\end{tabular}
\end{table}

\subsection{\label{sec:retrieval}Historical Analog Retrieval}

A practical advantage of any well-structured embedding is that it enables case-based reasoning. Given a query state, one can search the historical record for the geometrically nearest states and ask when the injector last occupied a comparable configuration \cite{aamodt1994cbr, johnson2021billion}. To avoid trivial self-matches, neighbors within a $\pm$12-hour window around the query are excluded.

\begin{figure}
\includegraphics[width=\columnwidth]{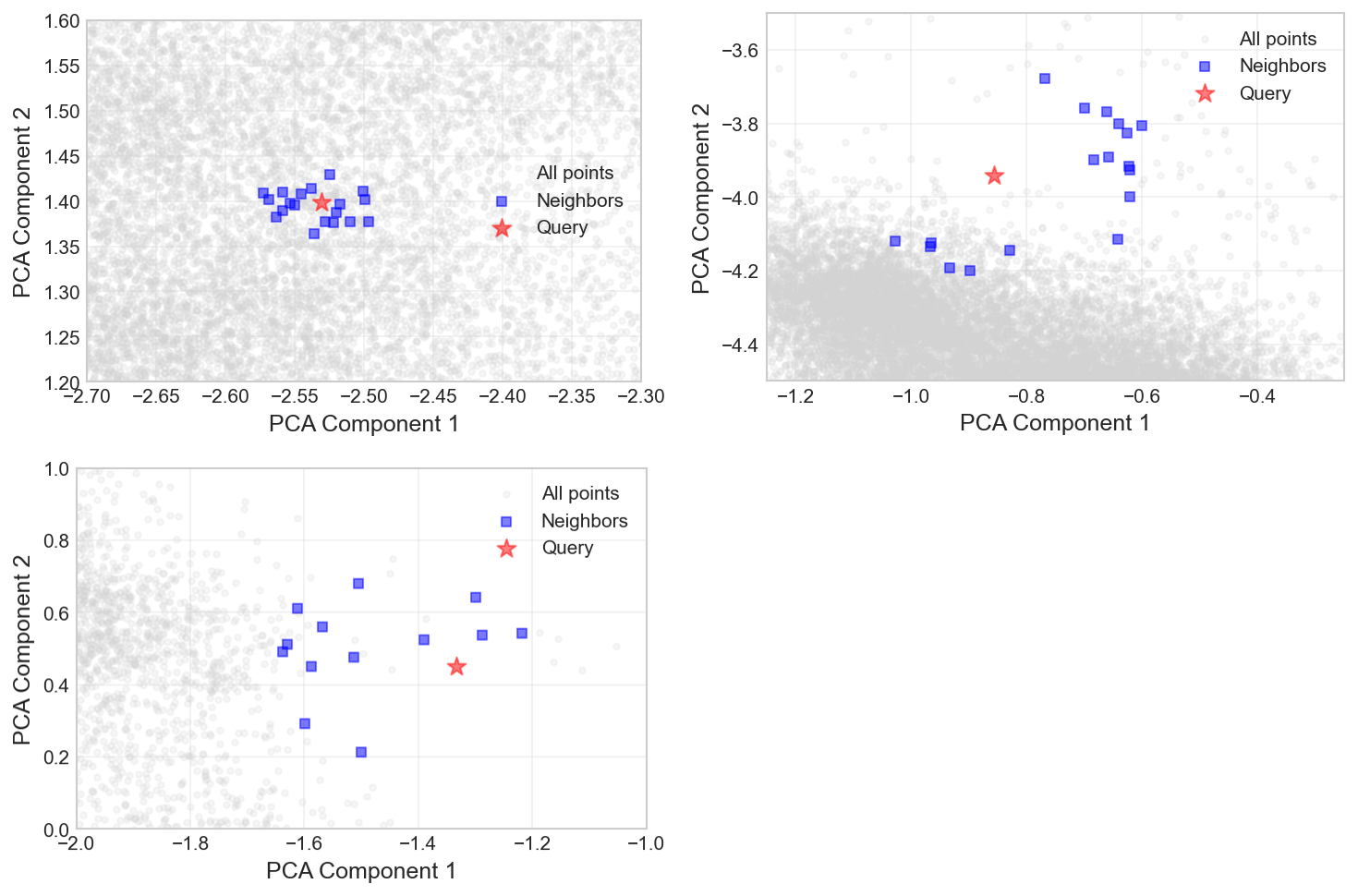}
\caption{\label{fig:retrieval}Historical-neighbor retrieval for three query 
types: a point from a large persistent cluster (upper left), a point near a 
large shift-scale transition (upper right), and a kNN outlier (lower left). 
Retrieval is performed in the full 16D embedding space while the plot shows 
the query and retrieved neighbors in a 2D PCA projection. Blue markers show 
retrieved neighbors after excluding the $\pm$12\mbox{-hour} window around 
the query; stars mark query locations.}
\end{figure}

Figure \ref{fig:retrieval} shows retrieval results for three query types. The neighbor search itself is performed in the full 16D embedding space; the PCA map is used only to visualize the query and its retrieved neighbors in two dimensions. A query drawn from the largest cluster (Cluster 9, July 12, 2022 at 20:26) returns 19 neighbors that remain tightly co-located with the query even in the 2D projection—the expected behavior for a highly repeatable operating state. A query near a large shift-change boundary (September 30, 2022 at 13:10) returns 18 neighbors that appear more spatially dispersed in the PCA view, consistent with that state lying near a transition corridor rather than a cluster core. An outlier query (September 27, 2022 at 16:58) returns 13 neighbors in a compact set, demonstrating that geometric outlier status does not mean the state is unique: it may be rare relative to the dominant modes but still recur with enough frequency to have historical precedent.

This case-based lookup is a qualitatively different operational capability from threshold-based alarming. It does not require a predefined definition of ``normal'' and instead leverages the full operational archive as a reference.

\section{\label{sec:interpretation}Physical Interpretation}

The fingerprint analysis (Table~\ref{tab:fingerprints}) and feature-variability ranking together connect the geometric structure of the embedding back to specific machine subsystems. Two complementary variability metrics provide different perspectives. The global mean-variance ranking—which asks which features fluctuate most on an hour-to-hour basis across the entire dataset—identifies superconducting cavity settings 0L02-7 and 0L02-8 as tied for the largest mean variance (2.40), followed by dipole MBF0L06 and quadrupole MQJ0L01 among settings. Reading variability is dominated by ion pumps VIP0L03E, VIP0L03D, and VIP0L03C. The stable-versus-unstable contrast—which asks which features expand most strongly during the hours when the embedding shows the largest internal spread—produces a consistent picture: 0L02-7 and 0L02-8 again rank first (variance increase 6.54), followed by dipoles MFD0I04A and MFD0I04, and corrector MQS0L01.

The recurrence of 0L02-7 and 0L02-8 in both rankings makes them especially compelling candidates for further investigation. These cavities are part of the quarter cryomodule (0L02), which sits early in the injector lattice and directly sets the initial beam energy. Their prominence in both routine variability and in the most dynamically evolving intervals suggests that changes to their gradients or phases are a primary driver of machine-state relocations across the dataset.

The cluster fingerprints extend this interpretation to specific regimes. As discussed in Section~\ref{sec:fingerprints}, Cluster~2 is organized by chopper cavity settings, indicating a mode defined by early beam formation, while Cluster~6 is organized by 0L04 SRF cavities, pointing to downstream configurations as the primary organizing variable. Cluster~9—the largest by point count, with a median dwell exceeding five days—is characterized by solenoid, capture, and corrector settings in the gun/low-energy region, suggesting a well-established, stable operating point in which the early capture optics are the primary differentiators.

These interpretations are suggestive, not definitive. Fingerprinting identifies correlates, not causes. But the consistency between the global variability rankings and the cluster-level differentiators strengthens the picture: the dominant directions of machine-state change in the embedding space correspond to physically interpretable subsystems.

\section{\label{sec:beamstudy}Controlled Beam-Study Validation}

To test whether the embedding also reflects deliberate, operator-imposed changes, we examine a controlled beam study in which the injector was reconfigured from one stable setup to another through 62 incremental changes. These changes were applied sequentially and spanned horizontal correctors, vertical correctors, quadrupoles, and—in one multi-element step—RF gradients, phases, and quadrupoles simultaneously.

\begin{figure}
\centering
\includegraphics[width=0.80\textwidth]{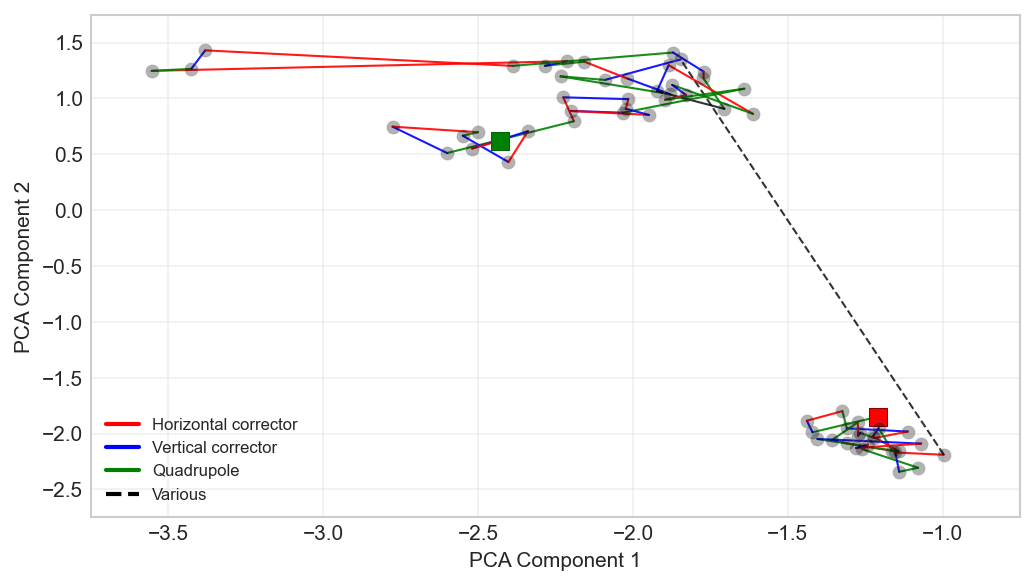}
\caption{\label{fig:beamstudy}PCA projection of the 62-step controlled beam 
study. Start and end configurations are marked by red and green squares, 
respectively. Edge colors denote the class of element changed; the dashed 
``various'' step involves simultaneous RF gradient, phase, and quadrupole changes 
and produces the largest single relocation. The coherent trajectory demonstrates 
that the embedding responds meaningfully to deliberate reconfiguration.}
\end{figure}

The PCA projection of this study is shown in Fig. \ref{fig:beamstudy}. Rather than appearing as disconnected scattered points, the sequence traces a smooth, interpretable trajectory through state space from the initial to the final configuration. Most steps produce local movement within a shared neighborhood: horizontal correctors, vertical correctors, and quadrupole changes all fall within a coherent region, consistent with their relatively limited influence on the global machine state. The single compound "various" step—involving simultaneous RF and magnet changes—produces a distinctly larger relocation between two separated regions.

This result serves as validation that the learned embedding is not merely organizing historical data into post-hoc structure: it behaves as a genuine operational coordinate system in which deliberate reconfiguration appears as structured motion rather than as noise.

\section{\label{sec:discussion}Discussion}

\subsection{\label{sec:add}What Embeddings Add Beyond Single-Channel Monitoring}

Traditional single-channel monitoring detects anomalies only when a specific channel crosses a predefined threshold. This approach is well suited to hardware faults and out-of-range readings, but it cannot detect operational conditions that are unusual only in the aggregate—where individual channels each appear normal but their combination is unusual relative to the machine's history. The embedding-based approach inverts this: it first asks whether the overall machine state is typical, and only then asks which channels are responsible for any detected deviation. The 99.6\% jitter compliance rate established in Section \ref{sec:stability} is not a result that could be obtained from any single channel; it reflects a reading-space dispersion computed across 267 simultaneous readbacks. Similarly, retrieval of historical analogs is inherently a whole-machine question. There is no single channel whose value uniquely specifies the injector's operational configuration, so case-based lookup requires an embedding that condenses the full state into a searchable representation.

\subsection{\label{sec:pca}Alternative Representations}

The GNN is not necessarily the only viable representation for organizing injector state space. Applying PCA directly to the standardized 393-dimensional raw machine features—without any trained model—still yields broad time-ordered islands that broadly resemble the GNN embedding map, although with less clean separation between regimes and less sharp transition structure. That result is important because it identifies a much simpler baseline workflow. The issue is not necessarily the runtime cost of a trained GNN, which can be modest at this scale, but the additional machinery required to build and sustain the representation: graph construction, model training, hyperparameter tuning, validation, and future maintenance as the machine evolves. Some applications may benefit sufficiently from direct feature-space analysis because it avoids that added pipeline. At the same time, the reduced separation in the raw-feature projection confirms that the contrastive GNN is doing meaningful work in organizing the state space.

Looking further ahead, the GNN represents one point on a spectrum of representation complexity. Transformer-based encoders \cite{vaswani2017attention} offer a middle ground: they can capture long-range dependencies among beamline elements without requiring an explicit graph structure, may generalize more gracefully across configuration changes, and have seen increasing adoption in scientific machine learning \cite{gorishniy2021revisiting}. Exploring such architectures as alternatives or complements to the GNN is a natural direction for future work, particularly as injector configurations evolve and the cost of maintaining a fixed graph representation grows.

\subsection{\label{sec:limits}Limitations}

Several limitations of the present study should be acknowledged. First, the GNN model is tied to a specific graph representation of the injector beamline. When the beamline configuration changes substantially—as it did during the scheduled maintenance period following March 2023, when elements were removed, added, and rearranged—the fixed-graph representation may no longer be valid, requiring retraining or fine-tuning before new data can be analyzed consistently with historical embeddings. This retraining burden is a practical constraint on the method's long-term applicability.

Second, the clustering and outlier results are sensitive, to some degree, to the choice of hyperparameters. The HDBSCAN minimum cluster size, the kNN neighbor count, and the merge gap for interval construction. The quantitative results reported here reflect a specific set of choices, and the cluster count, dwell statistics, and outlier counts would shift under different settings. We have aimed to report robustly supported qualitative conclusions—that regimes are persistent, that large relocations are rare, that outlier episodes are brief—that are not sensitive to small parameter variations.

Third, the jitter baseline was estimated from five anchor windows chosen for their operational stability. The observed window-to-window variation in p95 (from 5.52 on January 4, 2022 to 19.57 on June 13, 2022) indicates that the effective noise floor is not stationary across operating contexts. In the present analysis, all one-hour windows are normalized against a single pooled baseline. If that baseline were chosen more conservatively---for example, by normalizing against a larger reference value drawn from the upper end of the anchor-window distribution---the resulting stability ratios would be smaller, and fewer windows would be flagged as unstable. That would reduce sensitivity to modest departures from the nominal jitter floor. Conversely, a context-dependent baseline, in which the reference jitter level is conditioned on operating regime, beam-current range, or another state descriptor, would make the comparison more local and could improve specificity by reducing false positives caused by comparing dissimilar operating conditions to a single global reference. Determining how best to define such adaptive baselines is an important direction for future work.

Finally, the overlap analysis between noise dwells and kNN outliers demonstrated that the combined screen is sensitive not only to unusual machine states but also to data-quality failures—in this case, archiver gaps that caused multiple PVs to return undefined values. This is a useful property for a screening tool, but it underscores that manual follow-up is necessary before any flagged interval can be interpreted operationally. Data-quality filtering prior to embedding or anomaly scoring would reduce the noise in such screens.

\subsection{\label{sec:nextsteps}Path to Real-Time Deployment}

All analyses in this paper were performed retrospectively on a stationary historical dataset. The long-term goal is a real-time operational capability in which new machine snapshots are embedded continuously and the state-space view is updated live. In such a system, new embeddings could be generated every two minutes alongside updates to regime assignment, stability metrics, outlier flags, and the nearest-neighbor reference pool. The feasibility of this pipeline depends on whether the model can be updated incrementally as the machine configuration evolves. As noted above, lighter-weight representations—whether direct feature-space projections or transformer-based encoders—may offer a more maintainable path to continuous deployment, and benchmarking the operational tradeoffs among these approaches is an important next step.

\section{\label{sec:conclusion}Conclusion}

We have shown that GNN-based machine-state embeddings provide a practical operational coordinate system for the CEBAF injector. The 14-month operational history organizes into a small number of persistent, well-separated neighborhoods in a 16-dimensional learned space, with ten recurring regimes identified by density-based clustering and confirmed by persistence statistics. The dominant clusters are entered only a handful of times per year and maintain median dwells measured in days. Large relocations between neighborhoods are rare and episodic—detectable as isolated spikes in the shift-to-shift step-size distribution—while 99.6\% of one-hour operating windows fall within an empirically derived jitter baseline. Geometric outlier screening narrows the operational archive to a small, auditable set of intervals, and nearest-neighbor retrieval enables case-based reasoning over the historical record. A controlled beam study validates that the embedding responds meaningfully to deliberate reconfiguration. Together, these capabilities demonstrate that machine-state embeddings support holistic, data-driven operational monitoring in ways that single-channel inspection cannot achieve. The present work establishes an analysis framework; its full value will be realized when combined with real-time data streams and integrated with supporting information sources such as logbooks, downtime records, and operational documentation.

\section{\label{sec:acknowledgments}Acknowledgments}

This material is based on work supported by the U.S. Department of Energy, Office of Science, Office of Nuclear Physics under Contract No. DE-AC05-06OR23177.

\bibliographystyle{abbrvnat}
\bibliography{references}  

\section{\label{sec:supplement}Supplementary Methods: Logbook Analysis Workflow}

For the September 30, 2022 case study, logbook entries were analyzed to connect a large embedding-space relocation event to operator-recorded machine context. The goal of this workflow was not to automate root-cause attribution, but to assemble a chronologically ordered, human-authored narrative record that could be reviewed efficiently with the aid of a large language model.

Entries were retrieved via a Streamlit application that wraps the Jefferson Lab logbook REST API. For this study, the retrieval window spanned September 28--October 1, 2022, centered on the transition identified in the embedding analysis. Queries were restricted to entries of type ELOG and explicitly excluded AUTOLOG records. This distinction was important because AUTOLOG entries comprise the majority of logbook traffic but are machine generated and often contain little narrative context, whereas the purpose here was to recover operator descriptions, interventions, and observed symptoms.

Only the body text of each retained entry was used for downstream analysis. HTML markup was converted to plain text using BeautifulSoup, with paragraph and list structure preserved where possible to maintain readability. The raw collection was then filtered to retain entries likely to contain meaningful human-authored narrative. Two criteria were applied: the cleaned body text had to contain at least 50 words, and the fraction of recognizable English words had to exceed 0.70. These thresholds were chosen empirically to suppress entries consisting primarily of device names, numerical readbacks, or other low-context content, while preserving entries that described operational actions or observations.

The filtered results were normalized into JSON Lines (JSONL) format, with one record per entry. Each record contained a normalized ISO-8601 timestamp, the original timestamp string, the log number, the entry title, and the cleaned body text. Two lightweight enrichment steps were then added. First, heuristic flags were assigned by scanning the title and body for keywords associated with broad classes of operational activity, including configuration changes, fault or trip events, maintenance actions, procedural steps, and shift-summary content. These flags were intended only as cues for later interpretation; they were not treated as authoritative labels. Second, CEBAF-specific entities were extracted using regular expressions targeting machine-relevant identifiers such as cavity and zone names, device identifiers, and hall references. As with the flags, these extracted entities were used as contextual aids rather than as ground truth. The enriched records were sorted chronologically before being passed to the language model.

The resulting JSONL file was supplied to an OpenAI-compatible chat interface. The system prompt instructed the model to act as a CEBAF operations analyst, to ground its claims in the cleaned log text, to treat flags and extracted entities as heuristic aids rather than evidence, and to reason explicitly about temporal order. The prompt also directed the model to distinguish direct textual evidence from inference and to state clearly when the available record was insufficient to support a conclusion. In this work, the analysis was run using both Claude Haiku and GPT-4o.

This workflow has several limitations. Most importantly, the filtering thresholds are heuristic. Brief but consequential entries may be excluded if they fall below the word-count cutoff, while some retained entries may still contribute little interpretive value. In addition, keyword-based flags and regular-expression entity extraction can both miss relevant context and introduce false positives. The workflow should therefore be understood as a retrieval-and-organization aid for human review, not as a source of definitive operational labels.






\end{document}